18/06/03 ##################################### Latex

\documentclass[12pt]{article}

\begin{document}
\begin{flushright}
Stockholm\\ USITP 03-01\\18 juni 2003
\end{flushright}

\thispagestyle{empty}
\medskip
\begin{center}

{\large\bf Wess-Zumino actions and Dirichlet Boundary Conditions for Super $p$-branes with Exotic Fractions of Supersymmetry}
\\

\vspace{10mm}
I. Bengtsson$^{\rm a} $ and  A.A. Zheltukhin$^{\rm a,b}$ 
\end{center}
\begin{center}
$^{a}$ Institute of Theoretical Physics, University of Stockholm, SCFAB,\\
 SE-106 91 Stockholm, Sweden \\
$^{b}$ Kharkov Institute of Physics and Technology, 61108, Kharkov,  Ukraine
\vskip 15.mm
\end{center}
\begin{center}
{\bf Abstract}
\end{center}   
\vskip 5.mm

The general solutions in the  models of closed and open superstring and super $p$-branes with exotic fractions of the $N=1$ supersymmetry are considered and the spontaneously broken character of the $OSp(1,2M)$ symmetry of the models is established. It is shown that extending these models by Wess-Zumino terms generates the  Dirichlet boundary conditions for superstring and super $p$-branes. Using the generalized Wess-Zumino terms new  $OSp(1,2M)$ invariant super $p$-brane and $Dp$-brane-like actions preserving $\frac{M-1}{M}$ fraction of supersymmetry are proposed. For  $M=32$ these models suggest new superbrane vacua of $M$-theory preserving 31 from 32 global supersymmetries.

\vspace{10mm}

\section {Introduction}
Recently  new progress in the tracing of M-theory symmetries \cite{dufli}, \cite{hull} based on the development of the generalized holonomy conception \cite{dufstel} has been achieved.\footnote{Let us note that this conception permits an extension by  the lengthening of the spinor components of the connection $\Omega_{M}$. An example of the extension has been studied in \cite{tugzhe} for  $N=1,2$ supersymmetric electrodynamics, where the  covariant derivative $D_M$ 
lengthening  
$D_M \rightarrow \nabla_{M}= 
D_{M} + i\mu{\tilde W}_M$ with 
 $\tilde W_M=\frac{i}{4}(0,-\sigma_{\mu\alpha\dot\alpha} F^{\mu\dot\alpha},
 {\tilde\sigma}^{\mu\dot\alpha\alpha} F_{\mu\alpha})$
for  the $N=1$ spinor derivatives, and with  
$\tilde W_M=-\frac{i}{4}(0,D_{\alpha}^{i}W, {\bar D}^{\dot\alpha i}\bar W)$ for  the $N=2$ spinor derivatives, were considered. The spinor components of the connection $\tilde W_M$ take into account the anomalous magnetic moment (AMM) $\mu$ of charged and neutral particles with spin $1/2$ and generate the Pauli term. Taking into account of the AMM of $N=2$ massive superparticles is necessary to restore $\kappa-$symmetry in its interactions with  $N=2$ extended Maxwell supermultiplet \cite{uvaz}.} 
The generalized holonomy conception classifies vacuum states permitted by the centrally extended supersymmetry algebra \cite{agit},\cite{gght} and introduces  new  hidden space-time symmetries. It was shown in  \cite{hull} that the  holonomy extension in M-theory to the $SL(32,R)$ local symmetry is  necessary to include the fermionic degrees of freedom and to permit exotic vacuum states preserving 31 from 32 supersymmetries \cite{gah},\cite{bazil}. 

The string/brane  description of the vacuum state with the so high supersymmetry  was given by the model \cite{zuv1} of tensionless superstring and super $p$-brane. A connection of this model with the description \cite{bazil} (see also \cite{guz}) of the BPS states in M-theory was discussed in \cite{band}. 
 The model \cite{zuv1} develops the approach \cite{curt},\cite{bersez} and \cite{zli} to the description of string/brane  dynamics in superspaces extended by the addition of tensor central charge (TCC) coordinates. The centrally extended superspaces are characterized by the orthosymplectic symmetries and are closely connected with gauge theory of massless fields with higher spins \cite{frons}, \cite{vas} which  already appear in the quantized superparticle models with exotic supersymmetries \cite{balu},\cite{bls}. 

It was observed in \cite{west} that $OSp(1,64)$ symmetry is spontaneously broken in $D=11$ supergravity which is the low energy  phase of $M$/string-theory containing  massive higher spin states. This observation gave a reason to suppose that the superbrane microscopic structure also may be described in terms of the spontaneously broken orthosymplectic symmetries \cite{vas}.
 Taking into account the connection of tensionless strings and branes with higher spin field theory \cite{bo},\cite{witten} it is important to understand whether the $OSp(1,2M)$ symmetry of the model \cite{zuv1} is spontaneously broken.
 
Here we study this question for the case of closed and open tensionless superstring and super $p$-brane and find that the $OSp(1,2M)$ symmetry of the model is spontaneously broken by the general static solutions of the brane equations of motion. This effect is similar to the  partial supersymmetry  breaking by the super four-brane \cite{hulpol} and  the generalized  coordinates of the model \cite{zuv1} are the Goldstone fields of the $OSp(1,2M)$ symmetry. These Goldstone  fields may be associated with  effective long wave description of the  vacua in microscopic higher spin theories. 
 Also we construct  new topological Wess-Zumino like superstring and super  $p$-brane actions generating the Dirichlet boundary conditions and spontaneously breaking supersymmetry and $OSp(1,2M)$ symmetry. In addition  we propose a  new set of the $OSp(1,2M)$ invariant super $p$-brane  and $Dp$-brane like actions  preserving  $\frac{M-1}{M}$ fraction of the $N=1$ supersymmetry.

\section  {A simple super $p$-brane model with extra supersymmetry: the general solution and symmetries.}
 
The exactly solvable supersymmetric model of closed tensionless super $p$-brane\\ ($p=1,2,3,...)$ with extra $\kappa$-symmetry
\begin{equation}\label{1}
S_p=\frac{1}{2}\int d\tau d^{\,p}\sigma \rho^\mu (U_a W_\mu^{ab} U_b),
\end{equation} 
has been  studied in \cite{zuv1}. This model includes the Cartan differential one-form $W_{ab}$ 
\begin{equation}\label{2}
\quad W_{ab}=dY_{ab}-2i(d\theta_a\theta_b + d\theta_b\theta_a )
\end{equation}
invariant under the  $N=1$ global supersymmetry transformations 
\begin{equation}\label{3} 
\delta_\varepsilon\theta_{a}=\varepsilon_a ,\quad 
\delta_\varepsilon Y_{ab}=2i(\theta_a\varepsilon_b +\theta_b\varepsilon_a ),
\quad \delta_\varepsilon U_{a}=0
\end{equation} 
of the generalized superspace composed by the spin-tensor 
$Y_{ab}$, the Grassmannian Majorana spinor $\theta_a$ and an auxiliary commuting Majorana spinor $U_{a}$ \cite{volzh} parametrizing the light-like density of the brane momentum. The world-volume density $\rho^\mu=(\rho^\tau,\vec\rho) $ \cite{banzh}, invariant under the $N=1$ supersymmetry (\ref{3}), provides reparametrization invariance of $S_p$ (\ref{1}).
The real symmetric spin-tensor  $Y_{ab}$
\begin{equation}\label{4}
Y_{ab}\equiv x_{ab} + z_{ab}
\end{equation} 
 unifies the space-time coordinates $x_{m}$ and the TCC coordinates $z_{kl..m}$
\begin{equation}\label{5}
x_{ab}=x_{m}(\gamma^{m}\,C^{-1})_{ab},\quad 
z_{ab}=iz_{mn}(\gamma^{mn}\,C^{-1})_{ab}+z_{mnl}(\gamma^{mnl}\,C^{-1})_{ab}+...
\end{equation} 
of the $D$-dimensional Minkowski space-time with $D=2,3,4 \; mod(8)$. The
 spin-tensor  $Y_{ab}$ is a realization of the symmetric matrix of generalized symplectic coordinates previously considered in \cite{frons}, \cite{vas}. 
The action $S_p$ is invariant under the transformations of the enhanced $\kappa-$symmetry 
\begin{eqnarray}\label{6}
\delta_\kappa\theta_a=\kappa_a  ,\quad
\delta_\kappa Y_{ab}=-2i(\theta_a\kappa_b + \theta_b\kappa_a ), \nonumber\\ 
  \delta_\kappa U_{a}=0, \quad \delta_\kappa \rho^\mu=0,
\end{eqnarray} 
with the parameter $\kappa$ restricted by one real condition
\begin{equation}\label{7}
 U^{a}\kappa_{a}=0
\end{equation} 
and the super $p$-brane model (\ref{1}) preserves $\frac{M-1}{M}$ fraction of the $N=1$ supersymmetry, where $M$ is the dimension of the Majorana spinors  $\theta_a$ and $U_a$. 

 The action (\ref{1}) is  presented in the equivalent form \cite{zuv1}
\begin{equation}\label{8}
S_p=\frac{i}{2}\int d\tau d^{\,p}\sigma
\,\rho^\mu\{[(U^a\partial_\mu \tilde Y_a)
-(\partial_\mu U^a \tilde Y_a)] - 
\tilde\eta \partial_{\mu}\tilde\eta \},
\end{equation} 
where the Majorana spinor $\tilde Y_a$  is defined by the relation
\begin{equation}\label{9} 
 i\tilde Y_a= Y_{ab}U_{}^b - \tilde \eta \theta_a
\end{equation} 
and is a new effective variable substituted for $ Y_{ab}$ and $\tilde\eta$ 
 \begin{equation}\label{10}
\tilde\eta=-2i (U^{a}\theta_a)
\end{equation} 
is the Lorentz invariant Grassmannian field describing the Goldstone fermion of the model.
The action (\ref{8}) is the component representation of the $OSp(1,2M)$ 
invariant action  
 \begin{equation}\label{11} 
S_p=\frac{1}{2}\int d\tau d^{\,p}\sigma\,\rho^\mu 
\partial_{\mu}Y^{\Lambda} G_{\Lambda\Xi}Y^{\Xi},
\end{equation} 
 where  $Y^{\Lambda}=( iU^{a}, \tilde Y_a, \tilde\eta )$ is a real $OSp(1,2M)$ supertwistor and  $G_{\Lambda\Xi}= (-1)^{\Lambda\Xi+1}{}G_{\Xi\Lambda}$ is the invariant supersymplectic metric previously considered in superparticle dynamics \cite{balu}.
The equations of motion  following from $S_p$ (\ref{11})
\begin{eqnarray}\label{12}
2\rho^{\mu}\partial_{\mu}Y^{\Lambda}+\partial_{\mu}\rho^{\mu}Y^{\Lambda}=0,
\nonumber \\
\partial_{\tau}Y^{\Lambda} G_{\Lambda\Xi}Y^{\Xi}=0,\nonumber\\
\partial_{\vec\sigma}Y^{\Lambda}G_{\Lambda\Xi}Y^{\Xi}=0
\end{eqnarray}
 are invariant under the linearly realized  $OSp(1,2M)$ symmetry, worldvolume reparametrizations and the Weyl gauge symmetry  \cite{zuv2}
\begin{equation}\label{13}
\rho'^{\mu}=e^{-2\lambda(\tau,\vec\sigma)}\rho^\mu, \quad  
Y'^\Sigma=e^{\lambda(\tau,\vec\sigma)}Y^\Sigma.
\end{equation}

In the partially fixed reparametrization gauge  \cite{zuv1}
\begin{equation}\label{14}
\rho^{i}(\tau,\vec\sigma)=0, \quad (i=1,2,...,p),
\end{equation} 
removing  $p$ of $(p+1)$ components of the worldvolume density $\rho^\mu(\tau,\vec\sigma)$ without  breaking of the Weyl and  $OSp(1,2M)$ symmetries, the general solution of
Eqs.(\ref{12}) is given by
\begin{eqnarray}\label{15}
 Y^{\Lambda}(\tau,\vec\sigma)= 
\frac{1}{\sqrt{\rho^\tau(\tau,\vec\sigma)}}\,{\cal Y}^{\Lambda}(\vec\sigma),
\nonumber\\
\rho^{i}(\tau,\vec\sigma)=0, \quad (i=1,2,...,p).
\end{eqnarray} 
 The static  fields  ${\cal Y}^{\Lambda}(\vec\sigma)$ in (\ref{15}) are restricted by the $p$  initial data constraints
\begin{equation}\label{16}
\partial_{\vec\sigma}{\cal Y}^{\Lambda}(\vec\sigma)G_{\Lambda\Xi}{\cal Y}^{\Xi}(\vec\sigma)=0.
\end{equation} 
 which are the invariants of the Weyl and  $OSp(1,2M)$ symmetries.  In the case of closed super $p$-brane the components of ${\cal Y}^{\Lambda}(\vec\sigma)$ and  $\rho^\tau(\tau,\vec\sigma)$ are periodic functions of $\sigma^i$
 \begin{equation}\label{17}
 {\cal Y}^{\Lambda}(\sigma^{i} + 2\pi)= {\cal Y}^{\Lambda}(\sigma^i),
\quad \rho^\tau(\tau,\sigma^{i} + 2\pi)=\rho^\tau(\tau,\sigma^{i})
\end{equation} 
 The components  of the arbitrary supertwistor  ${\cal Y}^{\Lambda}(\vec\sigma)$ in the general solution (\ref{15}) are the invariants of the Weyl gauge  symmetry (\ref{13}) due to the presence of the $\rho^\tau(\tau,\vec\sigma)$ factor.  
However, they form the linear representation of the $OSp(1,2M)$ group, because $\rho^\tau$ is the invariant of this group.
The  $\rho^\tau(\tau,\vec\sigma)$-factor in (\ref{15}) concentrates all dependence of the general solution on the evolution parameter $\tau$ and it may be removed by the additional to (\ref{14}) gauge fixing
\begin{equation}\label{18}
\partial _\tau \rho^\tau(\tau,\vec\sigma)=0.
\end{equation}
The gauge condition (\ref{18}) breaks the Weyl symmetry, but  
 preserves the $OSp(1,2M)$ symmetry and simplifies the general solution  (\ref{15}) to  the pure static form 
\begin{eqnarray}\label{19}
Y^{\Lambda}(\tau,\sigma^i)=Y^{\Lambda}_{0}(\sigma^i),
\nonumber\\
\partial _\tau \rho^\tau(\tau,\vec\sigma)=0, \quad 
\rho^{i}(\tau,\vec\sigma)=0, \quad (i=1,2,...,p),
\end{eqnarray}
where $\rho^\tau(\tau,\vec\sigma)=\rho^\tau_0(\sigma )$ was moved in $Y^{\Lambda}_{0}(\sigma^i)$. One remarkss that the solutions (\ref{15}) and (\ref{19}) are equivalent on the classical level, because of a correlation between the Weyl and space-time conformal symmetries on the quantum level of the tensionless string treatment \cite{ilst}.

The components of the static field $Y^{\Lambda}_{0}(\sigma^i)$ describe the shape of the super $p$-brane. The superbrane has a freedom to choose any shape restricted by the initial data constraints (\ref{16}) and this shape will remains frozen during the evolution.  Any other shape  obtained from the initially  randomly chosen by any transformation belonging to the  $OSp(1,2M)$ group will have the same rights. However, a fixing of the  brane shape by any fixed  initial data for  $Y^{\Lambda}_{0}(\sigma^i)$ will break the $OSp(1,2M)$ symmetry.
From the point of view of the general theory of system with broken global symmetry \cite{dvd}  fixing of the form of $Y^{\Lambda}_{0}(\sigma^i)$ may be interpreated as a choice of the  vacuum state of the underlying field system with the spontaneously broken global $OSp(1,2M)$ symmetry. 
As a result, the static fields $Y^{\Lambda}_{0}(\sigma^i)$ are interpreted similarly to \cite{hulpol} as the Goldstone fields associated with the spontaneously broken $OSp(1,2M)$ symmetry and the action (\ref{11}) is an effective long wave action for the Goldstone fields associated with the super $p$-brane. It proves the spontaneously broken character of the $OSp(1,2M)$ symmetry as the symmetry of the brane action (\ref{11}).

Using the supersymmetry laws (\ref{3}) and the definitions (\ref{9}),(\ref{10}) of the components of the supertwistor $Y^{\Lambda}=(iU^{a},\tilde Y_a, \tilde\eta )$ we find the transformation properties of the Goldstone fields under the supersymmetry transformations from $OSp(1,2M)$  
\begin{equation}\label{20}
\delta_\varepsilon\tilde Y_{a}= 2i\tilde\eta\varepsilon_a,\quad 
\delta_\varepsilon\tilde\eta= -2iU^{a}\varepsilon_a , \quad 
 \delta_\varepsilon U_{a}=0, \quad  \delta_\varepsilon\rho^{\mu}=0.
\end{equation} 
The $N=1$ supersymmetry transformations (\ref{20}) are nonlinear, as it have to be for the spontaneously broken symmetries \cite{dvd}, because the original Goldstone fields  $\theta_a$ are presented in (\ref{20}) by only one their projection $(U^{a}\theta_a)$
. The absence of other $(M-1)$ projections of $\theta_a$ on $(M-1)$ basis spinors means the  disappearance  of $(M-1)$ Goldstone fermions corresponding to the unbroken fractions of the $N=1$ supersymmetry, because of the presence of the enhanced $\kappa$-symmetry (\ref{6}) restricted by the condition (\ref{7}) 
$$ U^{a}\kappa_{a}= 0.$$   
The non-zero projection $(U^{a}\varepsilon_a)$ of the supersymmetry parameter $\varepsilon_a$
\begin{equation}\label{21}
U^{a}\varepsilon_a\neq 0
\end{equation}
defines the direction of the spontaneously broken $(\frac{1}{M})$ fraction of the $N=1$ supersymmetry which can not be compensated any of the  $(M-1)$ $\kappa$-symmetry transformations. So, the condition (\ref{21}) is antipodal to the condition (\ref{7}) 
in the correspondence with the aforesaid and the Goldstone fermion $\tilde\eta$ has a non-zero shift (\ref{21}).

It is easy to check that the action $ S_{p}$ (\ref{8}), and respectively 
(\ref{11}), are invariant under the $N=1$ global supersymmetry transformations (\ref{20}), because of the cancellation between the contributions given by $\tilde Y_{a}$ and the fermionic Goldstone field $\tilde\eta$ 
\begin{equation}\label{22}
\delta_\varepsilon S_{p}= -\int d\tau d^{\,p}\sigma
\,\rho^\mu\{[U^a\partial_\mu \tilde\eta
-\partial_\mu U^a \tilde \eta] \varepsilon_a -[U^a \partial_\mu \tilde\eta -
\tilde\eta \partial_{\mu}U^a]\varepsilon_a\}=0.
\end{equation} 

An interesting and open question is to clarify the effect of the boundary terms for the dynamics of the open super $p$-branes and we turn to this question below.

\section {Boundary conditions for the open super $p$-brane} 

Here we study the case of open super $p$-brane (\ref{11}).
The contribution of the boundary terms in the variation of $S_p$ (\ref{11}) is given by  
\begin{equation}\label{23}
\delta {S_p}|_\Gamma= \oint ds_{\mu} \rho^\mu Y^{\Lambda} G_{\Lambda\Xi}\delta Y^{\Xi},
\end{equation}
 where $ds^{\nu}= \frac{1}{p!}\varepsilon^{\nu\mu_1\mu_2...\mu_p}dS_{\mu_1\mu_2...\mu_p}$.
 Here, we consider the variational problem with the fix initial ($\tau=\tau_i$) and final ($\tau=\tau_f$) data, so the integral along the super $p$-brane profile for $\tau=(\tau_i,\tau_f) $ does not contribute to $\delta {S_p}|_\Gamma$ (\ref{23})
\begin{equation}\label{24}
\int_{s_{\tau}} ds_{\tau} \rho^\tau Y^{\Lambda} G_{\Lambda\Xi}\delta Y^{\Xi}|_{\tau_i}^{\tau_f}=0.
\end{equation}
 As a result, the variation $\delta {S_p}|_\Gamma$ (\ref{23}) is filled  out by the integrals along the $p$-dimensional boundaries of the brane worldvolume 
containing the $\tau$-direction 
\begin{equation}\label{25}
\delta {S_p}|_\Gamma = \Sigma_{i=1}^{i=p}
\int_{s_i} ds_{i} \rho^{i}Y^{\Lambda}G_{\Lambda\Xi}\delta Y^{\Xi}
|_{\sigma^{i}=0}^{\sigma^{i}=\pi}.
\end{equation}
In the case of variational problem with free ends, i.e. when the field variations on the $p$-brane boundaries are arbitrary, the vanishing of these hypersurface terms in $\delta {S_p}|_\Gamma$  (\ref{25}) gives the open super $p$-brane
boundary conditions 
\begin{equation}\label{26}
\rho^{i}Y^{\Lambda}|_{\sigma^{i}=0,\pi} = 0, \quad (i=1,2,...,p).
\end{equation}
One of the solutions of (\ref{26}) is 
\begin{equation}\label{27}
 \rho^{i}(\tau,\vec\sigma)|_{\sigma^{i}=0,\pi}=0,\quad (i=1,2,...,p)
\end{equation}
 
The second possibility to satisfy the boundary conditions (\ref{26}) implies the zero  boundary conditions for the supertwistor $Y^{\Lambda}=(iU^{a},\tilde Y_a, \tilde\eta )$ values on the boundaries
\begin{equation}\label{28}
Y^{\Lambda}(\tau,\sigma)|_{\sigma^{i}=0,\pi}=0, \quad (i=1,2,...,p)
\end{equation}
or, equivalently, in terms of the  supertwistor components 
\begin{equation}\label{29}
U^{a}|_{\sigma^{i}=0,\,\pi}=0, \quad
\tilde Y_a|_{\sigma^{i}=0,\,\pi}=0, \quad
\tilde\eta|_{\sigma^{i}=0,\, \pi}=0,  \quad (i=1,2,...,p).
\end{equation} 
The boundary conditions (\ref{27})  for $\rho^{i}$ and  (\ref{28})  for $Y^{\Lambda}$  are invariant under the  Weyl symmetry (\ref{13}),  $N=1$ global supersymmetry (\ref{20}) and other homogenious transformations of $OSp(1,2M)$. 
The boundary conditions (\ref{27}) will be automatically satisfied in the invariant gauge (\ref{14})
$$
\rho^{i}(\tau,\vec\sigma)=0, \quad (i=1,2,...,p).
$$
 As a result, the general solution (\ref{15}) for the closed super $p$-brane in this gauge  gives  also the general solution of the boundary problem (\ref{12}),(\ref{27}) for the open super $p$-brane.

Concerning the boundary conditions (\ref{28}), one can note that the zero boundary values $U_{a}|_{\sigma^{i}=0,\pi}=0$ (\ref{29}) result in some problem in the geometric interpretation of the auxiliary spinor field $U_{a}$ as a basic constituent of the local spinor repere attached to the super $p$-brane worldvolume. For example, in the case of the $4$-dimensional Minkowski space, where $U^{a}$ is treated \cite{volzh} as one of the components of the Newman-Penrose dyads \cite{newpen}, these boundary conditions result in the condition
\begin{equation}\label{30}
(U^{a}(\tau,\vec\sigma)V_{a}(\tau,\vec\sigma))|_{\sigma^{i}
=0,\pi}=0, \quad (i=1,2,...,p)
\end{equation} 
which breaks the basis relation  $U^{a}V_{a}=1$ defining the dyads $U_{a},\, V_{a}$ \cite{newpen}. To preserve this condition the spinor field $V_{a}$ should be singular on the brane/string boundaries and it signals on some instabilities on the brane boundaries. 
Therefore, the solution (\ref{27}) have to be choosen for the considerd simple model (\ref{11}) and in this case the open and closed super $p$-brane are described by the same general solution (\ref{15}) for the static Goldstone fields.
This result is based on use of the gauge condition (\ref{14}) for the auxiliary  field $\rho^\mu$.

To overcome the problem of the singular character of the boundary conditions (\ref{28}) we need to extend the simple action (\ref{11}) and 
to this end we may generalize the topological actions studied in \cite{guz}, \cite{ilst}. An example of that generalization  will be done in the next section, where we will present of a topological action which yields the Dirichlet boundary conditions for open superstring,  resulting to the spontaneous breakdown of the $OSp(1,2M)$ symmetry and $N=1$ supersymmetry.

\section {A topological action generating the Dirichlet \\
boundary conditions for the superstring }

The superstring action with enhanced supersymmetry given by 
\begin{equation}\label{31} 
S_{WZ}=\frac{\beta}{2}\int _{\tau_i}^{\tau_f}\int_{0}^{\pi}G_{\Lambda\Xi}dY^{\Lambda}\wedge dY^{\Xi}
\end{equation}
contributes only on the superstring ends  and yields the Dirichlet boundary conditions similar to those for the   Nambu strings \cite{polch}.
 The integrand in the integral $S_{WZ}$ (\ref{31}) is a total derivative and 
 is  presented in the form of the integral along the one-dimensional boundary of the superstring worldsheet
\begin{equation}\label{32}
 S_{WZ}=-\frac{\beta}{2}\oint dY^{\Lambda} G_{\Lambda\Xi} Y^{\Xi}.
\end{equation}
The integral (\ref{31}) is similar to the curvature integral for the open string
\begin{equation}\label{33} 
S_{R}=
-\frac{C}{4\pi}\int_{\tau_i}^{\tau_f}\int_{0}^{\pi}R\sqrt{-g}\,d\tau d\sigma,
\end{equation} 
where $R/2$ is the Gauss curvature of the string worldsheet. It was shown in \cite{zh23} that taking into account of the nonlinear boundary conditions generated by $S_{R}$ reveals a topological structure of the string action extrema. To find the effect resulted in by  $ S_{WZ}$ (\ref{31}) one notes that the integrand of $S_{WZ}$ (\ref{32}) coincides with the differential form $(U_a W_\mu^{ab} U_b)$ in (\ref{1}) and therefore $S_{WZ}$ is invariant of the original symmetries of the action (\ref{11}) besides of the Weyl gauge symmetry (\ref{13}). The latter restriction follows 
from the absence of the $\rho^\mu$ density in the integral (\ref{32}) which results in it change
\begin{equation}\label{34} 
 S^\prime_{WZ}=-\frac{\beta}{2}\oint e^{2\lambda}dY^{\Lambda} G_{\Lambda\Xi} Y^{\Xi}
\end{equation}
under the Weyl transformation (\ref{11}). It means that the Weyl symmetry is explicitly broken by the boundary terms, already on the classical level unlike the Green-Schwarz superstring, where the breakdown  appears only on the quantum level.

 The variation of the Wess-Zumino term (\ref{32}) gives  
\begin{equation}\label{35} 
\delta S_{WZ}=-\beta\oint dY^{\Lambda}G_{\Lambda\Xi}\delta Y^{\Xi}-\frac{\beta}{2}\oint d(\delta Y^{\Lambda}G_{\Lambda\Xi} Y^{\Xi})=\beta\int _{\tau_i}^{\tau_f}
\partial_\tau Y^{\Lambda}G_{\Lambda\Xi}\delta Y^{\Xi}|_{\sigma=0}^{\sigma=\pi},
\end{equation} 
where the initial and final variational conditions $\delta Y^{\Xi}(\tau_i,\sigma)=0$,\, $\delta Y^{\Xi}(\tau_f,\sigma)=0$  have been used. 
Next,
taking  into account the freedom in the  variations $\delta Y^{\Lambda}(\tau_i,\sigma)|_{\sigma=0,\pi}$ on the string ends we obtain  the following boundary conditions
\begin{equation}\label{36} 
\partial_\tau Y^{\Lambda}(\tau,\sigma)|_{\sigma=0,\pi}=0,
\end{equation}
presented  in the component form  as 
\begin{equation}\label{37}
\partial_\tau U^{a}|_{\sigma =0,\,\pi}=0, \quad
\partial_\tau \tilde Y_a|_{\sigma =0,\,\pi}=0, \quad
\partial_\tau \tilde\eta|_{\sigma =0,\,\pi}=0.
\end{equation}
The boundary conditions (\ref{36}) and (\ref{37}) are the equations of motion of the string ends and they are invariant under the $OSp(1,2M)$ symmetry  and supersymmetry transformations, because of their global character.  
 However, the general solution of these equations 
\begin{equation}\label{38} 
 Y^{\Lambda}(\tau,\sigma)|_{\sigma=0}= {\cal A}^{\Lambda}, \quad
 Y^{\Lambda}(\tau,\sigma)|_{\sigma=\pi}= {\cal B}^{\Lambda}, 
\end{equation}
 which contains the integration constants  
${\cal A}^{\Lambda}$ and $ {\cal B}^{\Lambda}$, defined by the initial data 
\begin{equation}\label{39} 
{\cal A}^{\Lambda}\equiv(iU_{\cal A}^{a},\tilde Y_{{\cal A}a}, \tilde\eta_{\cal A}),\quad {\cal B}^{\Lambda}\equiv(iU_{\cal B}^{a},\tilde Y_{{\cal B}a}, \tilde\eta_{\cal B}),
\end{equation}
 defining the position of string ends in the symplectic superspace. 
The choice of different values for the constant supertwistors ${\cal A}^{\Lambda}$ and $ {\cal B}^{\Lambda}$ means the choice of different vacuum states breaking the $OSp(1,2M)$ symmetry.
 Note that ${\cal A}^{\Lambda}$ and $ {\cal B}^{\Lambda}$  have dimension $L^1$ and their choice define a length scale in the model fixing the scale of $\beta$ in  (\ref{31}).
Let us note  the particular solution of Eqs.(\ref{38}) fixed by the zero values
 of the Goldsone fermion on the string ends
\begin{equation}\label{40} 
 \tilde \eta_{\cal A}= 0, \quad \tilde \eta_{\cal B}=0.
\end{equation}
 The solution (\ref{40}) will partially preserve the supersymmetry  if the conditions 
\begin{equation}\label{41} 
 U_{\cal A}^{a}\varepsilon_a=0, \quad U_{\cal B}^{a}\varepsilon_a=0
\end{equation}
 for the projection $(U^{a}\varepsilon_a)$ on the superstring ends are satisfied, as it follows from the transformation rules (\ref{20}).
The conditions (\ref{41}) impose two real conditons for the supersymmetry parameters $\varepsilon_a$ resulting to the breaking of $(\frac{2}{M})$ fraction of $N=1$ supersymmetry  or, in the special  case
\begin{equation}\label{42} 
U_{\cal A}^{a}=U_{\cal B}^{a},
\end{equation} 
 to the breaking  only $(\frac {1}{M})$ fraction of $N=1$ supersymmetry. 

\section {The superstring model with the Wess-Zumino term}
 
Here we show that the addition of the Wess-Zumino term (\ref{31}) in the original action removes the problem of the singular character of the second solution
 (\ref{28}) of the boundary conditions (\ref{26}). 
The extended action  
 \begin{equation}\label{43} 
S= S_1 + S_{WZ}= \frac{1}{2}\int _{\tau_i}^{\tau_f}\int_{0}^{\pi} d\tau d\sigma\,\rho^\mu 
\partial_{\mu}Y^{\Lambda} G_{\Lambda\Xi}Y^{\Xi} + \frac{\beta}{2}\int _{\tau_i}^{\tau_f}\int_{0}^{\pi}G_{\Lambda\Xi}dY^{\Lambda}\wedge dY^{\Xi}
\end{equation} 
modifies the boundary conditions (\ref{26}) to the conditions 
\begin{equation}\label{44}
[\rho^{\sigma}Y^{\Lambda}+\beta\partial_\tau Y^{\Lambda}(\tau,\sigma)]|_{\sigma=0,\pi}=0.
\end{equation}
The conditions (\ref{44}) are invariant under the  $OSp(1,2M)$ symmetry similarly to (\ref{27}) and (\ref{28}) and  their general solution 
\begin{equation}\label{45}
 Y^{\Lambda}(\tau,\sigma)|_{\sigma=0}= 
e^{-\int _{\tau_i}^{\tau}\frac{\rho^\sigma(\tau,0)}{\beta}}
{\cal A}^{\Lambda}, \quad
 Y^{\Lambda}(\tau,\sigma)|_{\sigma=\pi}=e^{-\int _{\tau_i}^{\tau}\frac{\rho^\sigma(\tau,\pi)}{\beta}} {\cal B}^{\Lambda},
\end{equation}
 includes the arbitrary integration constants ${\cal A}^{\Lambda}$, $ {\cal B}^{\Lambda}$ similar to (\ref{39}). So, one can see that the boundary conditions (\ref{45}) are not singular when $\rho^{\sigma}|_{ 0,\pi}\neq 0$.
A fixing  of the constant ${\cal A}^{\Lambda}$ and $ {\cal B}^{\Lambda}$ means a vacuum state choice and shows the spontaneously broken character  of the  $OSp(1,2M)$ symmetry of the action (\ref{43}).

The action (\ref{43}) differs from the Wess-Zumino like action (\ref{31}) by the presence of the equations of motion (\ref{12}) having the general solution (\ref{15}) 
\begin{equation}\label{46}
 Y^{\Lambda}(\tau,\sigma)= 
\frac{1}{\sqrt{\rho^\tau(\tau,\sigma)}}\,Y^{\Lambda}_{0}(\sigma),\quad 
\rho^{\sigma}(\tau, \sigma)=0
\end{equation} 
 if the gauge (\ref{14}) (for  $p=1$) is choosen.
The substitution of (\ref{46}) in the boundary conditions (\ref{44}) with 
$\rho^{\sigma}=0$  results in the boundary conditions
\begin{equation}\label{47}
\partial_\tau\rho^{\tau}(\tau,\sigma)|_{\sigma=0,\pi}=0
\end{equation}
which are satisfied by the additional gauge fixing (\ref{18}) 
$$
\partial_\tau\rho^{\tau}(\tau,\sigma)=0.
$$
In this gauge the general solution (\ref{46}) coincides with the static solution (\ref{19}) describing  the above studied closed and opened superstrings
\begin{equation}\label{48}
 Y^{\Lambda}(\tau,\sigma)= Y^{\Lambda}_{0}(\sigma),
\end{equation}
but has the Dirichlet boundary conditions (\ref{38}). One notes that the initial data $Y^{\Lambda}_{0}(\sigma)$ \\(\ref{48}) are  restricted by the constraint (\ref{16})
 \begin{equation}\label{49}
Y^{\prime\Lambda}_{0}(\sigma)G_{\Lambda\Xi}Y^{\Xi}_{0}(\sigma)=0.
\end{equation} 
The matching (\ref{45}) and (\ref{48}) confirms that the integration constants ${\cal A}^{\Lambda}$,  ${\cal B}^{\Lambda}$ coincide with the $ Y^{\Lambda}_{0}(\sigma)$  values  taken on the string ends ${\sigma=0,\pi}$
\begin{equation}\label{50} 
{\cal A}^{\Lambda}\equiv Y^{\Lambda}_{0}(0), \quad
{\cal B}^{\Lambda}\equiv Y^{\Lambda}_{0}(\pi).
\end{equation}
We conclude  that the superstring action (\ref{43}) with  the Dirichlet boundary conditions (\ref{45}) describes a static BPS state with the spontaneously broken $OSp(1,2M)$ symmetry.

\section { Wess-Zumino actions of higher orders}

Using the  $OSp(1,2M)$ invariant character of the differential one-form 
 $Y^{\Lambda} G_{\Lambda\Xi} dY^{\Xi}$ and two-form 
 $dY^{\Lambda} G_{\Lambda\Xi} dY^{\Xi}$ one can construct more general  $OSp(1,2M)$ invariant super p-brane actions with enhanced supersymmetry.
 At first,  we note that  the closed  $2n$-differential form 
$\Omega_{2n}=(G_{\Lambda\Xi}dY^{\Lambda}\wedge dY^{\Xi})^n$ 
\begin{equation}\label{51} 
\Omega_{2n}=d\wedge\Omega_{(2n-1)}
\equiv
G_{\Lambda_1\Xi_1}dY^{\Lambda_1}\wedge dY^{\Xi_1}\wedge...\wedge 
G_{\Lambda_n\Xi_n}dY^{\Lambda_n}\wedge dY^{\Xi_n}
\end{equation}
 which is not equal to zero, because of the symplectic character of the supertwistor metric $G_{\Lambda\Xi}$,  
can be used to generate the Dirichlet boundary terms for the open super $p$-brane $(p=2n-1)$ described by the generalized action (\ref{43})  
 \begin{equation}\label{52} 
S= S_{2n-1} + 
\beta_{(2n-1)}\int_{M_{2n}}\Omega_{2n}.
\end{equation}
Similarly to the open superstring case (\ref{32}), the Wess-Zumino integral
in  (\ref{52}) is transformed to the integral along the $(2n-1)$-dimensional boundary  $M_{2n-1}$ of the super $(2n-1)$-brane worldvolume 
\begin{equation}\label{53}
 \int_{M_{2n}}\Omega_{2n}=\oint_{M_{2n-1}} 
G_{\Lambda_1\Xi_1}Y^{\Lambda_1}\wedge dY^{\Xi_1}\wedge...\wedge 
G_{\Lambda_n\Xi_n}dY^{\Lambda_n}\wedge dY^{\Xi_n}.
\end{equation}
 The sufficient conditions for the vanishing of the variations of the integral  (\ref{53}) with the  fix initial and final data are the conditions 
\begin{equation}\label{54} 
\partial_\tau Y^{\Lambda}(\tau,\sigma)|_{\sigma^{i}=0,\pi}=0,  \quad (i=1,2,...,2n-1)
\end{equation}
 generalizing the Dirichlet boundary condition  (\ref{36}). Therefore, in the gauge  (\ref{14}) and  (\ref{18}) this open super $p$-brane is described by the pure static solution 
\begin{equation}\label{55}
 Y^{\Lambda}(\tau,\sigmaî)= Y^{\Lambda}_{0}(\sigma^i),  \quad (i=1,2,...,2n-1)
\end{equation}
generalizing the superstring static solution (\ref{48}).
On the other hand the integrals  (\ref{53})
\begin{eqnarray}\label{56}
S_{(2n-2)}=  \beta_{(2n-2)}\int_{M_{2n-1}}\Omega_{2n-1}, \nonumber \\
\Omega_{2n-1}
\equiv  G_{\Lambda_1\Xi_1}Y^{\Lambda_1}dY^{\Xi_1}\wedge...\wedge 
G_{\Lambda_n\Xi_n}dY^{\Lambda_n}\wedge dY^{\Xi_n}
\end{eqnarray}
 can be considered as the $OSp(1,2M)$ invariant actions for the new models of super $p$-branes $(p=2n-2)$ with enhanced supersymmetry. For $n=1$ we get the known action \cite{balu} for superparticles, but for $n=2,3$ we find the new actions for the supermembrane 
\begin{eqnarray}\label{57}
S_{2}=\beta_{2}\int_{M_{3}}\Omega_{3} 
=\tilde\beta_{2}\int d\tau d^{\,2}\sigma\,\varepsilon^{\mu\nu\rho} 
Y^{\Lambda}\partial_{\mu}Y_{\Lambda}
 \partial_{\nu}Y^{\Xi}\partial_{\rho}Y_{\Xi},
\end{eqnarray}
 or a domain wall in the symplectic superspace, and for the super four-brane
\begin{eqnarray}\label{58}
S_{4}=\beta_{4}\int_{M_{5}}\Omega_{5}
=\tilde\beta_{4}\int d\tau d^{\,4}\sigma\,\varepsilon^{\mu\nu\rho\lambda\phi} 
Y^{\Lambda}\partial_{\mu}Y_{\Lambda}
 \partial_{\nu}Y^{\Xi}\partial_{\rho}Y_{\Xi} 
\partial_{\lambda }Y^{\Sigma}\partial_{\phi}Y_{\Sigma}.
\end{eqnarray}
We shall analyse these models in another place.
 
\section { The Weyl symmetry restoration for the  Wess-Zumino actions}

A characteristic feature of the proposed  Wess-Zumino actions  is the explicit breaking of the Weyl  gauge symmetry  (\ref{13}).
When the  Wess-Zumino terms  are considered as the boundary terms generating the Dirichlet  boundary conditions for the superstring  (\ref{36}) and super  $p$-branes (\ref{54}) the breaking of the Weyl symmetry is localized at the boundaries. It shows that the spontaneous breaking of the  
 $OSp(1,2M)$ symmetry on the boundaries is accompanied by the explicit breakdown of the Weyl gauge symmetry on the boundaries. Because the Dirichlet boundary conditions are associated with the  $Dp$-branes attached on their boundaries \cite{polch},  a question on the action of $Dp$-branes in the symplectic superspaces considered here appears. It implies  the correspondent generalization of the proposed  Wess-Zumino actions. One of the  posssible generalizations is rather natural and is based on the observation  that the Weyl invariance of the considered  Wess-Zumino actions  may be restored by the minimal lengthening of the differentials $d \rightarrow D=(d-A)$, where the worldvolume  one-form $A$ is the gauge field associated with the Weyl symmetry. The  covariant differentials  $DY^{\Sigma }$ are  homogeneously transformed under the Weyl symmetry transformations (\ref{13})
\begin{equation}\label{59}
  (DY^{\Sigma })'\equiv ((d-A) Y^{\Sigma })^\prime = e^\lambda DY^{\Sigma}, 
\quad A^\prime=A+d\lambda.
\end{equation}
 Then the generalized $OSp(1,2M)$ invariant  two and one-forms 
\begin{eqnarray}\label{60}
(e^\phi DY^{\Sigma} G_
{\Sigma\Xi}DY^{\Xi})^\prime=e^\phi DY^{\Sigma} G_{\Sigma\Xi} DY^{\Xi},
\nonumber \\
(e^\phi Y^{\Sigma} G_
{\Sigma\Xi}DY^{\Xi})^\prime=e^\phi Y^{\Sigma} G_{\Sigma\Xi} DY^{\Xi}
\end{eqnarray}
become the  invariants of the  Weyl symmetry also, where  the compensating scalar field $\phi$, with the  transformation low
\begin{equation}\label{61}
\phi^\prime=  \phi - 2\lambda,
\end{equation}
was introduced. 
Then  the closed  $2n$-differential form 
$\Omega_{2n}=(G_{\Lambda\Xi}dY^{\Lambda}\wedge dY^{\Xi})^n$ 
may be changed by the Weyl invariant  $2n$-differential form 
$\tilde \Omega_{2n}=(e^{\phi}G_{\Lambda\Xi}DY^{\Lambda}\wedge DY^{\Xi})^n$
\begin{equation}\label{62} 
\tilde\Omega_{2n}\equiv e^{n\phi}
G_{\Lambda_1\Xi_1}DY^{\Lambda_1}\wedge DY^{\Xi_1}\wedge...\wedge 
G_{\Lambda_n\Xi_n}DY^{\Lambda_n}\wedge DY^{\Xi_n}, 
\end{equation}
and $\Omega_{2n-1}$ by  $\tilde \Omega_{2n-1}$
\begin{equation}\label{63}
\tilde\Omega_{2n-1}\equiv e^{n\phi}
Y^{\Lambda_1}\wedge DY_{\Lambda_1}\wedge...\wedge 
DY^{\Lambda_n}\wedge DY_{\Lambda_n}. 
\end{equation}
As a result, the actions (\ref{53}) is  transformed to the new super $(2n-1)$-brane action
\begin{equation}\label{64}
\tilde S_{(2n-1)}= \beta_{(2n-1)}\int_{M_{2n}}\tilde\Omega_{2n}=\beta_{(2n-1)}
\int  e^{n\phi}
G_{\Lambda_1\Xi_1}DY^{\Lambda_1}\wedge DY^{\Xi_1}\wedge...\wedge 
G_{\Lambda_n\Xi_n}DY^{\Lambda_n}\wedge DY^{\Xi_n}
\end{equation}
invariant under the $OSp(1,2M)$ and Weyl symmetries. Respectively, the action 
\begin{equation}\label{65}
\tilde S_{(2n-2)}= \beta_{(2n-2)}\int_{M_{2n-1}}\tilde\Omega_{2n-1}=\beta_{(2n-2)}\int e^{n\phi}
Y^{\Lambda_1}\wedge DY_{\Lambda_1}\wedge...\wedge 
DY^{\Lambda_n}\wedge DY_{\Lambda_n}
\end{equation}
will describe a new  $OSp(1,2M)$ and Weyl invariant super $(2n-2)$-brane.\\
These actions may be presented in the $Dp$-brane like form, e.g.
\begin{equation}\label{66} 
\tilde S_p=\tilde\beta_{p}\int d\tau d^{\,p}\sigma \,e^{\frac{(p+1)}{2}\phi}
\sqrt 
{|det[(\partial_{\mu}- A_{\mu})Y^{\Lambda} G_{\Lambda\Xi}(\partial_{\nu}- A_{\nu})Y^{\Xi}]|}, \, (p=2n-1),
\end{equation}
where $\tilde\beta_{p}$ is  the $Dp$-brane tension.

\section {Conclusion}
We considered the general solutions of the equations of motion in the simple model of closed and open tensionless superstring and super $p$-branes and found that these static solutions spontaneouly break the $OSp(1,2M)$ symmetry and $N=1$ supersymmetry. 
Next, we generalized this model to the higher orders in the derivatives of the Goldstone fields and constructed the new Wess-Zumino like actions supposed to describe tensile super $p$-branes. These actions generate the Dirichlet boundary conditions which, in particular, break the Weyl gauge symmetry.
 The introduction of additional vector and scalar fields restores the Weyl symmetry and results in the Weyl and $OSp(1,2M)$ invariant $Dp$-brane like actions.
The open problem is to find supersymmetric Y-M field theories having the considered superbranes as vacuum states spontaneously breaking the $OSp(1,2M)$ symmetry. One can conjecture that these branes appear as supersymmetric solutions of $D=11$ supergravity \cite{dufli},\cite{hull}, where the $OSp(1,64)$ symmetry is also spontaneously broken \cite{west}. Then a connection between 
 the $R^{31}$ holonomy and space-time symmetries \cite{dufli},\cite{hull} with the local Abelian shifts of the space-time and TCC brane coordinates by the null multivectors  \cite{zuv2} may appear. We will study these problems in another place.

\section {Acknowledgements} 

 We thank M. Cederwall, F. Hassan, K. Narain  and D. Uvarov for helpful discussions. A.Z. thanks Fysikum at the Stockholm University for the kind hospitality. The work was partially supported by the grant of the Royal Swedish Academy of Sciences and Ukrainian SFFR project 02.07/276.


\begin{thebibliography}{99}
\bibitem{dufli}
M.J. Duff and J.M. Liu,
Hidden Spacetime Symmetries and Generalized Holonomy in M-theory, hep-th/0303140.
\bibitem{hull}
C.M. Hull, Holonomy and Symmetry in M-theory, hep-th/0305039.
\bibitem{dufstel}
M.J. Duff and K. Stelle,  Phys. Lett. B 253 (1991) 113;\\
M.J. Duff, M-theory an manifolds of $G_2$ holonomy: the  first twenty years, 
 hep-th/0201062.
\bibitem{tugzhe}
  A.A. Zheltukhin and V.V. Tugai, JETP Lett. 61 (1995) 541;\\
 Phys. Rev. D 54 (1996) 4160; 
On Extension of  Minimality Principle in Supersymmetric Electrodynamics,  hep-th/9706114.
\bibitem{uvaz}
 A. A. Zheltukhin and D.V. Uvarov, JETP Lett. 67 (1998) 888;\\
 Phys. Rev. D 61 (1999) 015004;\\
D.V. Uvarov, N=2 supersymmetric Yang-Mills theory and the superparticle:
  twistor transform and $\kappa$-symmetry, hep-th/0305051.
\bibitem{agit}
J.A. de Azcarraga, J.P. Gauntlett, J.M. Izquierdo and  P.K. Townsend,\\
 Phys. Rev. Lett. 63 (1989) 2443.
\bibitem{gght}
J.P. Gauntlett, G. Gibbons, C.M. Hull and P.K. Townsend, \\
Comm. Math. Phys. 216 (2001) 431.
\bibitem{gah}
J.P. Gauntlett and C.M. Hull, JHEP 01 (2000) 004.
\bibitem{bazil}
I. Bandos, J.A. de Azcarraga, J.M. Izquierdo and J. Lukierski,\\
Phys. Rev. Lett. 86 (2001) 4451. 
\bibitem{zuv1}
 A. A. Zheltukhin and D.V. Uvarov, Phys. Lett. B 545 (2002) 183;\\ JHEP 08 (2002) 008.
\bibitem{guz} 
O.E. Gusev and  A.A. Zheltukhin, JETP Lett. 64 (1996) 487.
\bibitem{band}
I.A. Bandos,  Phys. Lett. B 558 (2003) 197.
\bibitem{curt}
T. Curtright, Phys. Rev. Lett. 60 (1988) 393.
\bibitem{bersez}
E. Bergshoeff and E. Sezgin,
Phys. Lett. B 392 (1995) 256.
\bibitem{zli}
 A.A. Zheltukhin and  U. Lindstr\"om,  Nucl. Phys. (Proc. Suppl.) B102/101 (2001) 126; 
JHEP 01 (2002) 034.
\bibitem{frons} 
C. Fronsdal, Masslesss particles, orthosymplectic symmetry and another type of Kaluza-Klein theory, Preprint UCLA/85/TEP/10, in Essays on supersymmetry, \\Reidel, 1986 (Mathematical Physics Studies, v.8).
\bibitem{vas}
M.A. Vasiliev, Phys. Rev. D 66 (2002) 066006;\\ Russ. Phys. J. 45 (2002); Izv. Vuz. Fiz. 2002 N7 (2002) 23.
\bibitem{balu} 
I. Bandos and J. Lukierski, Mod. Phys. Lett. A 14 (1999) 1257.
\bibitem{bls}
I. Bandos, J. Lukierski and D. Sorokin, Phys. Rev. D61 (2000) 045002.
 \bibitem{west}
P.C. West, JHEP 08 (2000) 007.
\bibitem{bo}
P. Haggi-Mani and B. Sundborg, JHEP 04 (2000) 031;\\
B. Sundborg, Nucl. Phys. (Proc. Suppl.) B102/101 (2001) 113.
\bibitem{witten}
 E. Witten, unpublished (see http://theory.caltech.edu/jhs60/witten/1.html).
\bibitem{hulpol}
J. Huges and J. Polchinski,
 Nucl. Phys. B 278 (1986) 147;\\
J. Huges, J. Liu and J. Polchinski, Phys. Lett. B 180 (1986) 370.
\bibitem{volzh}
D.V. Volkov and A.A. Zheltukhin,
JETP Lett. 48 (1988) 63; \\Lett. Math. Phys. 17 (1989) 141. 
\bibitem{banzh}
 I.A. Bandos and A.A. Zheltukhin, Fortschr. Phys. 41 (1993) 619.
\bibitem{zuv2}
A. A. Zheltukhin and D.V. Uvarov, Phys. Lett. B 565 (2003) 229.
\bibitem{ilst}
J. Isberg,  U. Lindstr\"om  and  B. Sundborg, 
Phys. Lett. B293 (1992) 321;\\
J. Isberg,  U. Lindstr\"om,  B. Sundborg and G. Theodoridis,\\
Nucl. Phys. B411 (1994) 122.
\bibitem{dvd} 
D.V. Volkov, Sov. J. Part. Nucl. 4 (1973) 1.
\bibitem{newpen} 
R. Penrose and M.A.H. Mac Callum, Phys. Rep. 6 (1972) 241.
\bibitem{polch} 
J. Polchinski, TASI Lectures on D-branes, hep-th/9611050. 
\bibitem{zh23}
 A. A. Zheltukhin, Sov. J. Nucl. Phys. 34 (1981) 311;
Phys. Lett. B 116 (1982) 147.

\end{thebibliography}
\end{document}